\documentclass[aoas,nameyear,dvips]{arximspdf}
\usepackage{graphicx}

\doi{10.1214/10-AOAS425}
\volume{5}
\issue{2A}
\pubyear{2011}
\firstpage{725}
\lastpage{745}

\makeatletter
\newcommand{\eqref}[1]{(\ref{#1})}

\let\epsilon\varepsilon

  \let\sv@tabnotetext\tabnotetext
  \let\sv@tabnotemark@fmt\tabnotemark@fmt
   \long\def\legend#1{{\let\tabnote@indent\leavevmode\sv@tabnotetext[]{}{#1}}}

\makeatother

\begin{document}
\begin{frontmatter}

\title{Bayesian hierarchical modeling for signaling pathway inference
from single cell interventional~data\thanksref{TITL1}}
\runtitle{HMs for signaling pathway inference}
\thankstext{TITL1}{Supported in part by National Science Foundation
Grant DMS-07-14817,
National Heart, Lung, and Blood Institute Contract N01 HV28186,
National Institute of Drug Abuse Grant P30 DA018343, National
Institute of General Medical Sciences Grant R01 GM59507,
Yale University Biomedical High Performance Computing Center and NIH
Grant RR19895, which funded the instrumentation.}

\begin{aug}
\author[A]{\fnms{Ruiyan} \snm{Luo}\corref{}\ead[label=e1]{ruiyan.luo@yale.edu}}
\and
\author[A]{\fnms{Hongyu} \snm{Zhao}\ead[label=e2]{hongyu.zhao@yale.edu}}
\runauthor{R. Luo and H. Zhao}
\affiliation{Yale University}
\address[A]{Department of Epidemiology and Public Health\\
Yale University School of Medicine\\
New Haven, Connecticut 06520\\
USA\\
\printead{e1}\\
\hphantom{E-mail:\ }\printead*{e2}} 
\end{aug}

\received{\smonth{9} \syear{2009}}
\revised{\smonth{9} \syear{2010}}

%
\begin{abstract}
Recent technological advances have made it possible to simultaneously
measure multiple protein activities at the single cell level.
With such data collected under different stimulatory or inhibitory
conditions, it is possible to infer the causal relationships
among proteins from single cell interventional data.
In this article we propose a~Bayesian hierarchical modeling framework
to infer the signaling pathway based on the posterior distributions
of parameters in the model.
Under this framework, we consider network sparsity and model the
existence of an association between two proteins both at the overall level
across all experiments and at each individual experimental level.
This allows us to infer the pairs of proteins that are associated with
each other and their causal relationships.
We also explicitly consider both intrinsic noise and measurement error.
Markov chain Monte Carlo is implemented for statistical inference.
We demonstrate that this hierarchical modeling can effectively pool information
from different interventional experiments through simulation studies
and real data analysis.
\end{abstract}

%
\begin{keyword}
\kwd{Bayesian network}
\kwd{dependency network}
\kwd{Gaussian graphical model}
\kwd{hierarchical model}
\kwd{interventional data}
\kwd{Markov chain Monte Carlo}
\kwd{mixture distribution}
\kwd{single cell measurements}
\kwd{signaling pathway}.
\end{keyword}

\end{frontmatter}

\setcounter{footnote}{1}

\section{Introduction}

Cells respond to internal and external changes through signaling networks.
One major research area in biology is to identify
signaling proteins and understand how they coordinate to
function properly. With recent technological advances in genomics
and proteomics, researchers now can monitor and quantify molecular
activities at the genome level, making it possible to reconstruct
signaling pathways from these high-throughput data. Although
efforts have been made to use microarray gene expression data and sequence
data to reveal signaling pathways [e.g., \citet{Liu-Ringner-2007}],
these data are limited in two important aspects. First,
signaling pathways function at the protein level, so measured gene
expression levels from microarrays at most can provide a proxy to the protein
activity levels. Second, each cell may behave differently from other
cells due to complex interactions among many proteins, some substantially.
Therefore, population level data collected by microarrays can mask
individual cell differences, making it difficult to infer
underlying pathways. In contrast, single cell
level protein activity data offer much richer information for pathway inference.

Flow cytometry [\citet{flow02}; \citet{nolan02}] is a~%
powerful fluorescence-based technology that can make rapid,
sensitive, and quantitative measurements of multiple proteins for
thousands of individual cells.
It can measure both a specific protein's expression level and protein
modification states
such as phosphorylation.
Therefore, phospho-protein responses to environmental
stimulations can be monitored at the single cell level for thousands of
cells very
efficiently, and this technology has been employed to infer signaling
pathways through gathering activity
levels of multiple proteins under different stimulatory or
inhibitory conditions [\citet{sachs-etal-05}]. We focus on the
analysis of single cell flow cytometry data in this
article.

Several methods have been applied for network inference based on
genomics data, including Bayesian Networks (BNs) [\citet{peer01};
\citet{peer05}], Markov Networks (MNs, also called Markov random
fields) [Wei and Li (\citeyear{li07}, \citeyear{li08})], and Dependency
Networks (DNs) [\citet{heckerman-etal-00}].
Common to all these methods, each protein (or gene) is represented by a node
and a dependency between two proteins is represented by an edge in the network.
More formally, we define a graph $\mathcal{G} = (V,E)$ with its nodes
$V=\{1,\ldots, P\}$ and an edge set $E$.
We use $X_i$ to refer to the value of the $i$th node, that is, the
expression level of the $i$th protein.
The methods differ in how the edges are inferred from the observed data.
In BN, the network is a directed acyclic graph where the state of each
node only depends on its immediate ancestors.
This structure imposes Markovian dependency among all the nodes stating
that each variable is conditionally independent of its nondescendants
given its parent variables.
So the joint likelihood for all the nodes, that is, proteins, can be
factored into a product of conditional probabilities.
BNs pose significant computational challenges to learn the network
structure because the model space to be explored is super-exponential
in the number of genes to be studied.
More recently, \citet{ellis-wong} proposed a method to reduce the bias
in the fast mixing algorithm proposed by \citet{friedman-koller} to
sample the BN structures from the posterior distribution.
MNs are undirected graphical models and are similar to BNs in
representation of dependencies: each random variable is conditionally
independent of all other variables given its neighbors.
Gaussian Graphical Models (GGMs) [\citet{lauritzen}; \citet
{schafer-strimmer05}; \citet{dobra-etal04}], a subclass of MNs,
assume a multivariate normal distribution as the joint distribution of
random variables.
The existence of an undirected edge in a GGM is implied by the nonzero
partial correlation coefficient derived from the precision matrix.
Some studies have found that BNs outperform GGMs in inferring networks
based on interventional data where the biological system
is perturbed through designed experiments, but GGMs may perform better
for observational data, for example, \citet{werhli-etal-06}.
DNs aim to reduce computational burden where a large number of genes
are modeled by building a collection of conditional distributions separately.
DNs define the conditional distributions $\{p(X_i|X_{-i})\}$ separately
for each $X_i$.
When we focus on sparse normal models, DNs
define a set of $P$ separate conditional linear regression models in
which $X_i$ is regressed
on a small selected subset of predictor variables, which are determined
separately.

Because a statistical association between two variables only implies
association not causation, standard DN and GGM approaches cannot be
used to
infer causal networks, a goal in signaling pathway analysis. In this
article we develop a Bayesian hierarchical modeling approach based
on DNs to address this limitation.
This is achieved through appropriate intervention experiments to
dissect directional influences.
To accommodate varying relationships among proteins under different
experimental conditions,
we allow a different set of regression models for each condition.
At the same time, the hierarchical framework imposes similar functional
forms across conditions to borrow information from different experiments.
As for causal inference, the basic idea is that for any protein $i$,
its regulators exert similar effects if it is not intervened,
and would have no effect when $i$ is controlled.
In contrast to standard regression models where the predictors are
assumed to be error-free,
our model allows measurement errors in predictor variables.

A large part of the difficulty in the standard BN computation is due to
the requirement that the
network be acyclic. Our approach is not guaranteed to give
acyclic networks. However, in terms of sensitivity and
specificity to detect true edges, our method is competitive
with the best methods that impose the acyclic graph assumption.
This is illustrated by our results in the example of \citet
{sachs-etal-05}.

The paper is organized as follows. In Section \ref{sec2} we describe two
hierarchical models (a general hierarchical model where no constraints
are imposed and a restricted hierarchical model where a symmetry
constraint is imposed) and the methods for statistical inference of the
network. For comparison, we also describe a nonhierarchical model where
all the experiments are pooled together for analysis.
Then we investigate the performance of these methods on simulated data
in Section \ref{sec3}.
In Section \ref{sec4} we apply these methods to data from a study of the
signaling networks of human primary naive CD4$^+$ cells [\citet
{sachs-etal-05}].
We finish the paper with discussions in Section \ref{sec5}.

\section{Methods}\label{sec2}
The primary goal of our statistical model is to infer causal influences
among proteins from interventional data.
In this section we describe three models: a hierarchical model (HM), a
restricted hierarchical model (RHM), and a nonhierarchical model (NHM),
that can be used to infer the relationships among proteins.
We also discuss statistical methods to infer causal networks in this section.

\subsection{Hierarchical model (HM)}\label{sec2.1}
First we discuss a Bayesian hierarchical model to infer the
relationships among
proteins both at the overall level across all experiments and under individual
experimental conditions. Our model incorporates both measurement
errors and the intrinsic noises due to the biological process and
unmodeled biological variations.

Let $P$ denote the number of proteins, $K$ denote the number of
experimental conditions, and $N_k$ denote the number of samples
(individual cells) under the $k$th condition, $k=1,2,\dots,K$.
We further let $\tilde{x}_{\mathit{ink}}$ denote the true activity level of the
$i$th protein in the $n$th cell under the $k$th experimental condition,
and $x_{\mathit{ink}}$ the measured value of its activity,
where $x_{\mathit{ink}} = \tilde{x}_{\mathit{ink}} + \epsilon^M_{\mathit{ink}}$,
and the measurement error $\epsilon^M_{\mathit{ink}}$ is a normal random
variable with mean 0 and standard deviation $\sigma^M$, that is,
$\epsilon^M_{\mathit{ink}} \sim\mathrm{N}(0,(\sigma^M)^2)$.
Our model assumes that there exists a linear relationship among the
activity levels of proteins.
That is, for each protein $i=1,2,\dots,P$ and for each condition
$k=1,2,\dots,K$,
%
\begin{equation}\label{xtilde.eq}
\tilde{x}_{\mathit{ink}}=\alpha_{i0}^{(k)}+\sum_{j\ne i}{\alpha
_{ij}^{(k)}\tilde{x}_{\mathit{jnk}}}+\epsilon_{\mathit{ink}}^I ,
\end{equation}
where $\epsilon_{\mathit{ink}}^I$ is the intrinsic noise and has a normal
distribution $\mathrm{N}(0,(\sigma^I_i)^2)$.\footnote{Here a
constant variance $(\sigma_i^I)^2$ is assumed for the intrinsic noises
of a particular protein. We can relax this assumption and allow varying
variances $(\sigma_{ik}^I)^2$ for intrinsic noises under different
experimental conditions. This extended model and simulation results are
described in Supplementary Material S1 [\citet{supp}].}
We assume that the error terms $\{\epsilon_{\mathit{ink}}^I\}$ are independent,
and are independent of the measurement errors $\{\epsilon^M_{\mathit{ink}}\}$.
In equation \eqref{xtilde.eq}, $\alpha_{ij}^{(k)}=0$ if
there is no linear relationship between the activity levels of
proteins $i$ and $j$ under the $k$th experimental condition.
A nonzero value of $\alpha_{ij}^{(k)}$ implies the existence of a
linear relationship (but not necessarily a causal effect).
To correctly infer the network among these proteins, we need to
first find, for each protein, the subset of proteins that are linearly
associated with its expression level, which is implied by
the set of nonzero coefficients in \eqref{xtilde.eq}.
The linear relationship among the true expression values
$\tilde{x}_{\mathit{ink}}$ implies that the observed values are also linearly
related:
%
\begin{eqnarray}\label{obsLine.eq}
x_{\mathit{ink}}&=&\alpha_{i0}^{(k)}+\sum_{j\ne i}{\alpha
_{ij}^{(k)}(x_{\mathit{jnk}}-\epsilon^M_{\mathit{jnk}})}
+\epsilon^I_{\mathit{ink}}+\epsilon_{\mathit{ink}}^M
\nonumber\\[-8pt]\\[-8pt]
&=& \alpha_{i0}^{(k)}+\sum_{j\ne i}{\alpha_{ij}^{(k)}x_{\mathit{jnk}}}
+\epsilon^I_{\mathit{ink}}+\epsilon_{\mathit{ink}}^M-\sum_{j\ne i}\alpha
_{ij}^{(k)}\epsilon^M_{\mathit{jnk}} .\nonumber
\end{eqnarray}
Comparing \eqref{xtilde.eq} and \eqref{obsLine.eq}, we can see
that correctly inferring the relationship in the network depends on the correct
inference of the set of nonzero coefficients in
\eqref{obsLine.eq}.

For each protein, we utilize indicator variables $z_{ij}=0$/1 to denote
the relationship between proteins $i$ and $j$ such that
$z_{ij}=1$ if and only if the coefficient of the $j$th protein in the
regression model for the $i$th protein is nonzero.
The values of $z_{ij}$ may differ under different experimental conditions.
For example, if protein $j$ regulates protein $i$, $z_{ij} = 1$ when
$i$ is not controlled and the association strength between the two
proteins should be similar under such conditions.
However, $z_{ij}=0$ when $i$ is controlled because the relation between
$X_i$ and $X_j$ is destroyed.
Therefore, it is natural to use~a~hie\-rarchical structure to formalize
this thinking.
We use $w_{ij}^{(k)}$ to denote the probability that $z_{ij} = 1$, that
is, protein $j$ is related to protein $i$ under the $k$th experimental
condition.
The prior we take for the regression coefficient~$\alpha_{ij}^{(k)}$
is a mixture of two distributions.
One is a point mass at zero, indicating that the $j$th protein is not
linearly related to the
$i$th protein under the $k$th condition.
The other is a normal distribution for nonzero
effects, with weight~$w_{ij}^{(k)}$.
Specifically, the prior for the slope coefficient $\alpha_{ij}^{(k)}$
($j \ne i$) is
%
\begin{equation}\label{alphaCondPrior.eq}
\alpha_{ij}^{(k)}|w_{ij}^{(k)},\alpha_{ij},\sigma^{\alpha}_{ij}
\sim\bigl(1-w_{ij}^{(k)}\bigr)\delta_0\bigl(\alpha_{ij}^{(k)}\bigr) +
w_{ij}^{(k)}\mathrm{N}\bigl(\alpha_{ij}^{(k)}|\alpha_{ij},(\sigma
^{\alpha}_{ij})^2\bigr) ,
\end{equation}
where $\delta_0(\cdot)$ indicates a point-mass at zero, and
$w_{ij}^{(k)}$ is the probability that $\alpha_{ij}^{(k)} \ne0$.
When $\alpha_{ij}^{(k)} \ne0$, the prior for $\alpha_{ij}^{(k)}$ is
$\mathrm{N}(\alpha_{ij},(\sigma^{\alpha}_{ij})^2)$ with a common
mean and
a common standard deviation across different experimental conditions.
Under this setup, information is shared for coefficients $\alpha
_{ij}^{(k)}$ across different conditions.
Similarly, we borrow information for $w_{ij}^{(k)}$ across different
experimental conditions by applying a beta
distribution as a prior for~$w_{ij}^{(k)}$ with a common mean
$w_{ij}$ and a common variance $w_{ij}(1-w_{ij})/(v_{ij}+1)$:
%
\begin{equation}\label{wCond0.eq}
w_{ij}^{(k)} | w_{ij},v_{ij} \sim\operatorname{Beta}\bigl(w_{ij}v_{ij},
(1-w_{ij})v_{ij}\bigr) .
\end{equation}
So $w_{ij}^{(k)}$ measures the probability that there is an association
between proteins $i$ and $j$ under the $k$th experimental condition,
and $w_{ij}$ measures the overall-level probability that the two
proteins are associated.

To complete the model,
we specify a beta distribution $\operatorname{Beta}(\beta_1,\beta
_2)$ for $w_{ij}$, a
normal distribution $\mathrm{N}(a^{(i)},\tau^{(i)})$ for $\alpha
_{ij}$ and gamma
distributions: $\mathrm{G}(\gamma_1,\gamma_2)$, $\mathrm{G}(\gamma
_3,\gamma_4)$ and
$\mathrm{G}(\gamma_5,\gamma_6)$ for $(\sigma_i^I)^{-2}$,
$(\sigma^{\alpha}_{ij})^{-2}$ and $(\sigma^M)^{-2}$, as their
respective prior distributions.
In the simulation studies, we take $\gamma_i=\beta_j=1$ for $i=1,2$
and $j=1,2,\ldots,6$, $a^{(i)}=0$ and $\tau^{(i)}=1000$ for
$i=1,\ldots,P$.
In real data analysis, we vary the hyperparameter values to study the
sensitivity of the inference results to these values.
We note that the posterior distribution is proper since we take proper
priors for all the parameters.

One attractive feature of this model is that when $w_{ij}^{(k)}$ is
integrated out,
the marginal distribution of $\alpha_{ij}^{(k)}$ is independent of $v_{ij}$:
%
\begin{equation}
\alpha_{ij}^{(k)}|w_{ij},\alpha_{ij},\sigma^{\alpha}_{ij} \sim
(1-w_{ij})\delta_0\bigl(\alpha_{ij}^{(k)}\bigr) +
w_{ij}\mathrm{N}\bigl(\alpha_{ij}^{(k)}|\alpha_{ij},(\sigma^{\alpha
}_{ij})^2\bigr) .
\end{equation}
Given $\alpha_{ij}^{(k)}$ and $w_{ij}$, when
$v_{ij}$ is specified, the posterior distribution of $w_{ij}^{(k)}$~is
%
\begin{equation}\label{wCond1.eq}
w_{ij}^{(k)}|w_{ij},v_{ij},\alpha_{ij}^{(k)}
\sim\operatorname{Beta} \bigl(w_{ij}v_{ij}+I\bigl(\alpha_{ij}^{(k)}\ne0\bigr),
(1-w_{ij})v_{ij}+I\bigl(\alpha_{ij}^{(k)}=0\bigr) \bigr) .\hspace*{-25pt}
\end{equation}
Hence, we can first sample the posterior distributions of
$\alpha_{ij}^{(k)}$ and $w_{ij}$, and then sample $w^{(k)}_{ij}$
according to equation \eqref{wCond1.eq}.

Under this model, the inference of the causal network consists of two steps.
First, based on the posterior means $\hat{w}_{(i,j)}$ of the
overall-level probability $0.5 \times(w_{ij}+w_{ji})$, we infer
whether there is an association between proteins~$i$ and $j$ with a
certain threshold $u_1$.
Second, for each pair of proteins $(i,j)$ that are inferred to be associated,
we determine their regulatory direction based on the experiment-level
probabilities $w^{(k)}_{ij}$ to infer the causal network.
The underlying assumption of our inference is that for a pair of
proteins ($i$~and~$j$) that has a regulatory relation, say, $i$
regulates $j$ ($i \rightarrow j$), controlling (inhibiting or
activating) over protein $j$ affects the activity of $j$ but not $i$,
resulting in much reduced or lack of association between $i$ and $j$;
controlling over protein $i$ affects the activity of $i$ and hence $j$,
keeping the association between them.
The posterior distributions of $w^{(k)}_{ij}$ given $\alpha
_{ij}^{(k)}$ and $w_{ij}$, with~$v_{ij}$ prespecified, is given in
equation \eqref{wCond1.eq}.
To better reflect the changes of~$w^{(k)}_{ij}$ for different $k$, we
use $v_{ij}\equiv0.1$ in our analysis, because larger values of~%
$v_{ij}$ (e.g., 10) are not able to reveal the changes in
$w^{(k)}_{ij}$, as the parameters in~\eqref{wCond1.eq} are dominated
by $v_{ij}$ when $v_{ij}$ is large, and $I(\alpha^{(k)}_{ij} \ne0)$
plays a smaller role in \eqref{wCond1.eq}.
To put this into more concrete terms, we consider $w_{ij}=0.9$ as an
example, which gives a strong support for the association between
proteins~$i$ and $j$.
The difference between the distributions $\operatorname
{Beta}(0.9\times10+1,0.1\times10)$ and $\operatorname
{Beta}(0.9\times10,0.1\times10+1)$ when $v_{ij}=10$ is much less than
that between $\operatorname{Beta}(0.9\times0.1+1,0.1\times0.1)$ and
$\operatorname{Beta}(0.9\times0.1,0.1\times0.1+1)$ when $v_{ij}=0.1$.

To determine the directions of edges, we calculate the posterior means
$\hat{w}^{(k)}_{ij}$ of $w^{(k)}_{ij}$ for all $k$ and $(i,j)$ pairs.
For each pair $(i,j)$, if all the values in a~stream (e.g., $\{\hat
{w}^{(k)}_{ji}\}_k$) are small (less than a threshold $u_2>0$, so the
signal in this stream is weak compared to noises), then we ignore this
stream and infer the causal relations only based on the other one ($\{
\hat{w}^{(k)}_{ij}\}_k$).
The inference is based on checking whether $\hat{w}^{(k)}_{ij}$ under
specific conditions decreases greatly compared to the highest value.
Let $S_{i,j}=\{k: k\in\{i,j\}\}$ denote the set of conditions under
which $i$ or $j$ is perturbed, and $|S_{i,j}|$ be its cardinality.
We propose the following four criteria to determine the causal
relationship between an associated protein pair ($i$, $j$):
%
\begin{table}[b]
\tablewidth=335pt
\caption{A summary of the nine experimental
conditions for the data in
Sachs et~al. (\protect\citeyear{sachs-etal-05})}\label{condition}%
\vspace*{-3pt}
\begin{tabular*}{\tablewidth}{@{\extracolsep{\fill}}lcc@{}}
\hline
& \textbf{Stimulus} & \textbf{Effect}\\
\hline
1 & CD3, CD28 & general perturbation\\
2 & ICAM2 & general perturbation\\[3pt]
3 & Akt-inhibitor & Inhibits Akt\\
4 & G0076 & Inhibits Pkc\\
5 & Psi & Inhibits Pip2\\
6 & U0126 & Inhibits Mek\\
7 & Ly & Inhibits Akt\\[3pt]
8 & PMA & Activates Pkc\\
9 & $\beta_2c\mathit{AMP}$ & Activates Pka\\
\hline
\end{tabular*}
\vspace*{-3pt}
\end{table}

\begin{figure}

\includegraphics{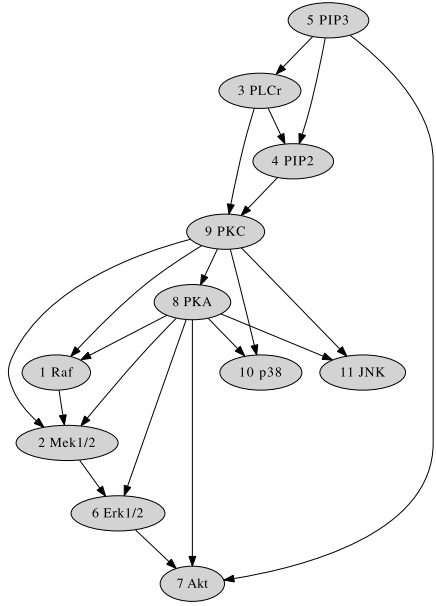}
\vspace*{-4pt}
%
\caption{Pathway adapted from Sachs et~al. (\protect\citeyear{sachs-etal-05}) by including
three missed edges and correcting one reversed edge. Nodes represent
proteins, and directed edges represent signal transduction.}\label{pathway.fig}
\vspace*{-6pt}
\end{figure}

\begin{itemize}
\item
Case 1: $|S_{i,j}|=1$, that is, $i$ or $j$ is only perturbed in one
condition. Without loss of generality, we suppose that protein $i$ is
controlled under condition~$k'$ ($S_{i,j}=\{k'\}$). If $\max_k\{\hat
{w}^{(k)}_{ij}\} - \hat{w}^{(k')}_{ij} > u_3$ for a threshold $u_3>0$,
then from stream $\{\hat{w}^{(k)}_{ij}\}_k$ we infer $j \rightarrow
i$. Otherwise, we infer $i \rightarrow j$. Similarly, we make an
inference from the stream $\{\hat{w}^{(k)}_{ji}\}_k$. If the
directions inferred from both streams are the same, say, $i \rightarrow
j$, we infer that direction as the direction of the edge between $i$
and $j$: $i \rightarrow j$. If the directions from both streams are
different, we say that the direction of the edge is undetermined.
Taking the conditions in Table \ref{condition} for the network in
Figure~\ref{pathway.fig} as an example, pairs $(1,2)$, $(1,8)$,
$(2,6)$ $(3,4)$, $(4,5)$, $(6,8)$, $(8,10)$, and $(8,11)$ belong to
Case 1.
\item
Case 2: $|S_{i,j}|>1$ and for all $k \in S_{i,j}$, the same protein,
say, $i$, is controlled. For each stream, for example, $\{\hat
{w}^{(k)}_{ij}\}_k$, if $\max_k\{\hat{w}^{(k)}_{ij}\} - \hat
{w}^{(k')}_{ij} > u_3$ for all $k' \in S_{i,j}$, then we infer $j
\rightarrow i$;
if $\max_k\{\hat{w}^{(k)}_{ij}\} - \hat{w}^{(k')}_{ij} \le u_3$ for
all $k' \in S_{i,j}$, then we infer $i \rightarrow j$;
otherwise, we do not infer a direction from this stream.
If both streams lead to a directional inference and the directions are
the same (Figure~\ref{wcond-v.fig}, top panel), or if only one stream
provides a directional inference, then we infer the direction of the edge.
Otherwise, the direction is undetermined.
For the conditions in Table \ref{condition}, pairs $(1,9)$, $(3,9)$,
$(5,7)$ $(6,7)$, $(9,10)$, $(9,11)$ belong to Case~2.
\item
Case 3: $|S_{i,j}|>1$ and both proteins are controlled in the
experiments. Let $S_{i,j}^{(i)}$ denote the set of conditions under
which protein $i$ is controlled, and $S_{i,j}^{(j)}$ the set of
conditions under which protein $j$ is controlled. For each stream, for
example, $\{\hat{w}^{(k)}_{ij}\}_k$, we calculate the differences of
$\hat{w}^{(k)}_{ij}$ when $i$ or $j$ is controlled:
$d^{(k_1k_2)}_{ij}=\hat{w}^{(k_1)}_{ij}-\hat{w}^{(k_2)}_{ij}$ for
each $k_1 \in S_{i,j}^{(i)}$ and $k_2 \in S_{i,j}^{(j)}$. If
$d^{(k_1k_2)}_{ij} > u_3$ for all $k_1 \in S_{i,j}^{(i)}$ and $k_2 \in
S_{i,j}^{(j)}$, 
we infer that $i \rightarrow j$; if $d^{(k_1k_2)}_{ij} \le-u_3$ for
all $k_1 \in S_{i,j}^{(i)}$ and $k_2 \in S_{i,j}^{(j)}$, we infer that
$j \rightarrow i$; otherwise, the direction is undetermined from this
stream. If both streams lead to a~directional inference and the
directions are the same, or if only one stream provides a directional
inference, then we infer the direction of the edge.
Otherwise, the direction is undetermined.
For the conditions in~Ta\-ble~\ref{condition}, pairs $(2,9)$, $(4,9)$,
$(7,8)$ $(8,9)$ belong to Case 3.
\item
Case 4: $|S_{i,j}|=0$, that is, no perturbation is conducted on either
protein. In this case, we cannot infer the causal relation.
\end{itemize}

The choices of the thresholds $u_1$, $u_2$, and $u_3$ will be discussed
in simulation studies. Generally speaking, the two steps involved in
causal network inference are based on the posterior distributions of
$w_{ij}$ and $w_{ij}^{(k)}$.
The overall-level probability $w_{ij}$ measures the strength
of the linear relationship between two proteins across all the
conditions. Based on
$w_{ij}$, we infer the set of proteins that are related from which we
determine an undirected graph.
The changes in the experiment-level probabilities $w_{ij}^{(k)}$ offer
insights on the directions of causal regulations.

\subsection{Restricted hierarchical model (RHM)}\label{sec2.2}
The $w_{ij}$ and $w_{ij}^{(k)}$ in \eqref{alphaCondPrior.eq}, \eqref
{wCond0.eq}, and \eqref{wCond1.eq} denote the probability that protein
$j$ is included in the linear model to predict the activity level of
protein $i$ across all the experiments and under the $k$th specific
condition, respectively.
In this framework, we may impose the constraint that $w_{ji}=w_{ij}$
and $w_{ji}^{(k)}=w_{ij}^{(k)}$ for each $k$, that is, the existence of
a linear relationship between proteins $i$ and $j$ is independent of
which variable is the predictor and which is the response variable.
We can infer the posterior distributions of $w_{ij}$ and $w_{ij}^{(k)}$
under this constraint, and call this model a restricted hierarchical
model (RHM).
Based on the posterior means of $w_{ij}$, we can infer whether proteins
$i$ and $j$ are associated with each other by setting up an appropriate
threshold $u'_1$.
For each associated pair, we can infer the causal relationship
according to the changes in $w_{ij}^{(k)}$.
The choice of the threshold will be illustrated in Section 3.1 and
Supplementary Material S3 [\citet{supp}] details the criteria in
determining the causal relations for the associated pairs of proteins.
Different from HM, we must prespecify $v_{ij}$ in RHM to sample from
the posterior distributions of~$w_{ij}$ and $w_{ij}^{(k)}$.
We will show how different values of $v_{ij}$ affect the network
inference in the following discussion.

\subsection{Nonhierarchical model (NHM)}\label{sec2.3}
To demonstrate the usefulness of the hierarchical model approach, we
also consider a
nonhierarchical model (NHM) as a reference model for comparisons.
The NHM assumes a linear model among the activity levels of
proteins and incorporates both measurement errors and intrinsic
noises as in equation \eqref{obsLine.eq}.
The main difference is that this NHM assumes identical regression
coefficients across different experimental conditions:
%
\begin{equation}
x_{\mathit{ink}}=\alpha_{i0}+\sum_{j\ne i}{\alpha_{ij}x_{\mathit{jnk}}}+\epsilon
_{\mathit{ink}}^{I}+\epsilon_{\mathit{ink}}^M-\sum_{j\ne i}\alpha_{ij}\epsilon
^M_{\mathit{jnk}} ,
\label{lineRegNonHier.eq}
\end{equation}
where the intrinsic noise $\epsilon_{\mathit{ink}}^I$ follows the normal distribution
$\mathrm{N}(0,(\sigma_i^I)^2)$, the measurement error $\epsilon
_{\mathit{ink}}^M$ follows the normal distribution $\mathrm{N}(0,(\sigma
^M)^2)$, and they are assumed to be independent.
As in HM, we also apply mixture distributions as priors for the
coefficients $\alpha_{ij}$:
%
\begin{equation}
\alpha_{ij} \sim(1-w_{ij})\delta_0(\alpha_{ij}) +
w_{ij}\mathrm{N}(\alpha_{ij}|a,\tau^2) .
\label{alpha.eq}
\end{equation}
The posterior distributions of $w_{ij}$ provide information about
whether proteins $i$ and $j$ are associated.
However, it is impossible to make causal inference from this model.

For all three models, we use MCMC methods to sample the posterior
distributions. Supplementary Material S2 [\citet{supp}] provides
details of the MCMC updates for HM. The MCMC updates for RHM and NHM
are similar and not shown in this paper.

\section{Simulation study}\label{sec3}

We first apply our methods to simulated data to illustrate how to infer
the causal network from the posterior distributions of the
overall-level probabilities $w_{ij}$ and the experiment-level
probabilities~$w_{ij}^{(k)}$, for both HM and RHM.
We also study how the inference differs between these two methods and
for different choices of $v_{ij}$.
We then study the performance of our methods on simulated data with
heavy tail distributed intrinsic noises.

We simulate data based on the network shown in Figure \ref
{pathway.fig}, which is adapted from \citet{sachs-etal-05} by
correcting one reversed edge and including three missed edges.
From Figure \ref{pathway.fig}, we can derive the parent set for each
node (protein).
For any protein $i$, we first generate the association strength $\alpha
_{ij}$ from the
uniform distribution over the interval $[0.5, 2]$, and randomly assign
the sign of $\alpha_{ij}$.
Given the activities of its parents, we simulate the activity $\tilde
{x}_i$ of protein $i$ from the normal
distribution: $\tilde{x}_i \sim\mathrm{N}(\alpha_{i0}+\sum_j
\alpha_{ij}\tilde{x}_j,(\sigma^I_i)^2)$,
where the sum extends over all parents of protein $i$.
Thus, we get the empirical distribution of $\tilde{x}_i$ when protein
$i$ is not intervened.
Let $x_i$ denote the observed expression level of protein $i$, then
$x_i$ is simulated from $\mathrm{N}(\tilde{x}_i, (\sigma^M)^2)$.
We simulate the interventional data as follows.
For an intervention experiment,
if the $i$th protein is inhibited, we sample $\tilde{x}_i$ from the
left tail of
its empirical distribution obtained when protein $i$ is not perturbed,
beyond the 5th percentile.
If the $i$th node is stimulated, we sample~$\tilde{x}_i$ from the
right tail of
the empirical distribution, beyond the 95th percentile.
We simulate a total of nine stimulatory or inhibitory interventional conditions,
as summarized in Table \ref{condition}.
Under each perturbation condition, we simulate expression levels for
each of the 11 proteins for 600 individual cells.
We consider two cases: (1) constant intrinsic variances $(\sigma
^I_i)^2\equiv1$ and (2) variable intrinsic variances with $\sigma^I_i
= 0.1 \times\sqrt{\operatorname{IG}(2,1)}$, where $\operatorname
{IG}(2,1)$ represents the inverse gamma distribution with mean 1 and
variance $\infty$.
Finally, we simulate data where the intrinsic noises are sampled from a
heavy tail distribution: $t(1)$, which represent a central $t$
distribution with one degree of freedom.

\subsection{\texorpdfstring{Constant intrinsic variance: $(\sigma^I_i)^2\equiv1$}%
{Constant intrinsic variance:}}
\label{sec3.1}
\subsubsection{Inference from HM}\label{sec3.1.1}
\begin{figure}

\includegraphics{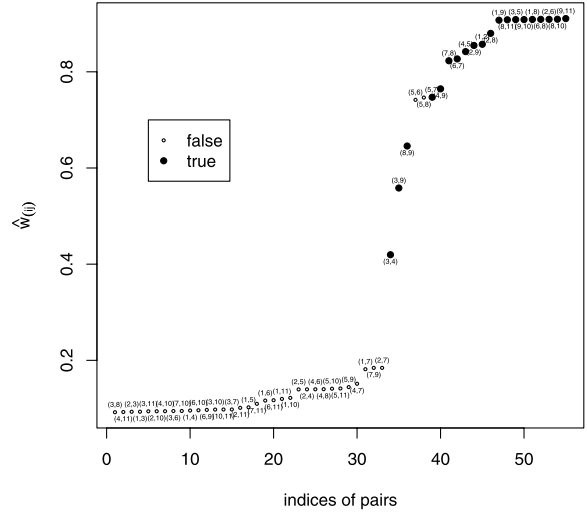}

\caption{Posterior means $\hat{w}_{(i,j)}$ of $(w_{ij}+w_{ji})/2$,
sorted in increasing order, from one MCMC run of the HM on the
simulated data with constant intrinsic variances. Large solid and small
empty circles represent true and false associations,
respectively.}\label{wMean.fig}
\end{figure}

Based on the simulated data, we obtain samples for both $w_{ij}$ and
$w_{ji}$ from their posterior distributions under HM.
To infer whether an association exists between proteins $i$ and $j$, we
obtain the posterior means $\hat{w}_{(i,j)}$ of the average of the
probability that each is included in the regression model of the other:
$(w_{ij}+w_{ji})/2$.
Higher values of $\hat{w}_{(i,j)}$ imply stronger evidence of
association between the two proteins.
Figure \ref{wMean.fig} shows the posterior means $\hat{w}_{(i,j)}$,
from one MCMC run, for each pair $(i, j)$ ($i<j$), in the ascending
order of $\hat{w}_{(i,j)}$.
Large solid circles represent true associations, and small empty ones
represent false ones.
We see that the true associations dominate the higher values of $\hat
{w}_{(i,j)}$.

\begin{figure}[b]

\includegraphics{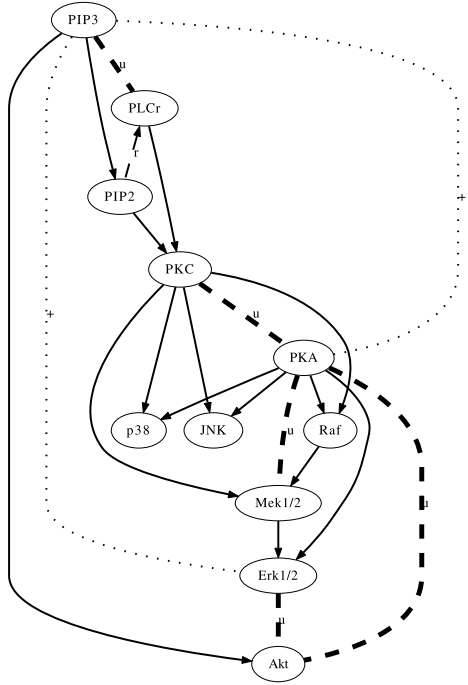}

\caption{Networks inferred by choosing associations with $u_1 \in
(0.3,0.4)$, $u_2=0.1$, $u_3 \in(0.3,0.5)$, and $u_f \ge0.8$ in five
MCMC runs of the HM on the simulated data with constant $\sigma_i^{I}$.
Solid arrowed lines represent correctly inferred true edges, dashed
thick lines with labels ``u'' represent edges whose directions cannot
be determined from the simulations, dashed arrowed thin lines with
labels ``r'' represent reversed edges, and dotted lines with labels
``+'' represent false positive edges.}\label{sml18-network.fig}
\end{figure}

To infer the pair of proteins that are associated, we need to set a
threshold~$u_1$ on the posterior means $\hat{w}_{(i,j)}$ so that those
above the threshold are inferred to be associated.
The permutation study\footnote{We permute the observations for each
protein and then analyze the permuted data with HM. The obtained
posterior means $\hat{w}_{(i,j)}$ are less than 0.1 for all $(i,j)$
pairs.} offers an over-liberal threshold ($<$0.1), based on which we
get over 40 associations with false positive rate $\ge$0.5.
Noting the jumps in the plot of $\hat{w}_{(i,j)}$, we propose to
choose the threshold where a big jump occurs.
Setting the threshold $u_1$ as any value between 0.2 and 0.4 and
choosing the pairs with $\hat{w}_{(i,j)}>u_1$, we get 22 associations
with 2 false positives.
When we have multiple MCMC runs, which lead to multiple plots of $\hat
{w}_{(i,j)}$, we can combine the inferences from them.
Figure S1 in the Supplementary Material [\citet{supp}] draws the
plots of $\hat{w}_{(i,j)}$ from four additional MCMC runs.
They show the same features as seen in Figure \ref{wMean.fig} that
true associations tend to have high~$\hat{w}_{(i,j)}$ values and jumps
exist in these plots.
These five MCMC runs lead to quite similar results: from four of them
we get 22 associations with 2 false positives, and from a fifth run we
get 21 associations with 2 false positives and~1 missing association,
when we choose $u_1$ between 0.3 and 0.4.
Let $u_f$ be the relative frequency that each association is selected.
When $u_1 \in(0.3,0.4)$ and $u_f \ge4/5$, we get 22 associations with
2 false positives (Figure \ref{sml18-network.fig}).

\begin{figure}

\includegraphics{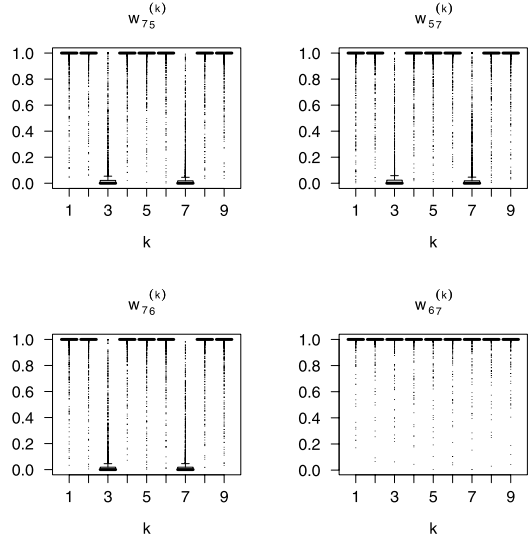}

\caption{Boxplots of the samples from the posterior distributions
of $w^{(k)}_{57}$ and $w^{(k)}_{75}$ \textup{(top panel)}, $w^{(k)}_{67}$ and
$w^{(k)}_{76}$ \textup{(bottom panel)} when $v_{ij}\equiv0.1$ for all $i$ and
$j$. This is from one MCMC run of the HM on the simulated data with
constant $\sigma_i^{I}$.}\label{wcond-v.fig}
\end{figure}

For the pairs of proteins that are inferred to be associated, we then
infer their causal directions based on the criteria listed in Section \ref{sec2.1}.
To better illustrate the criteria, we give two examples in Figure \ref
{wcond-v.fig}.
The top panel draws the boxplots of the samples from the posterior
distributions of $w^{(k)}_{57}$ and $w^{(k)}_{75}$.
Both show that the experimental-level probabilities greatly decreased
under conditions 3 and 7 where protein 7 (Akt) is inhibited (here $\max
_k\{\hat{w}^{(k)}_{57}\} - \hat{w}^{(k')}_{57} \ge0.8$ and $\max_k\{
\hat{w}^{(k)}_{75}\} - \hat{w}^{(k')}_{75} = 0.9$ for $k'=3,7$).
So we infer the direction $5 \rightarrow7$ (i.e., PIP3 $\rightarrow$ Akt).
The bottom panel tells a different story.
The posterior means of $w^{(k)}_{76}$ when $k=3$ or 7 are much smaller
than others ($\max_k\{\hat{w}^{(k)}_{76}\} - \hat{w}^{(k')}_{76} =
0.9$), indicating the causal relation $6 \rightarrow7$, but
$w^{(k)}_{67}$ keeps the same level under all conditions ($\max_k\{
\hat{w}^{(k)}_{67}\} - \hat{w}^{(k')}_{67} = 0$), indicating the
causal relation $7 \rightarrow6$.
The contradictory results from $w^{(k)}_{67}$ and $w^{(k)}_{76}$ lead
to the failure in determining the causal relationship between proteins
6 (Erk) and 7.
Taking $u_2=0.1$ and $u_3 \in(0.3,0.5)$, we infer a causal network as
shown in Figure \ref{sml18-network.fig}, which contains 14 true
directed edges, 5 edges whose directions are undetermined, 1 reversed
edge, and 2 false edges.

\begin{figure}

\includegraphics{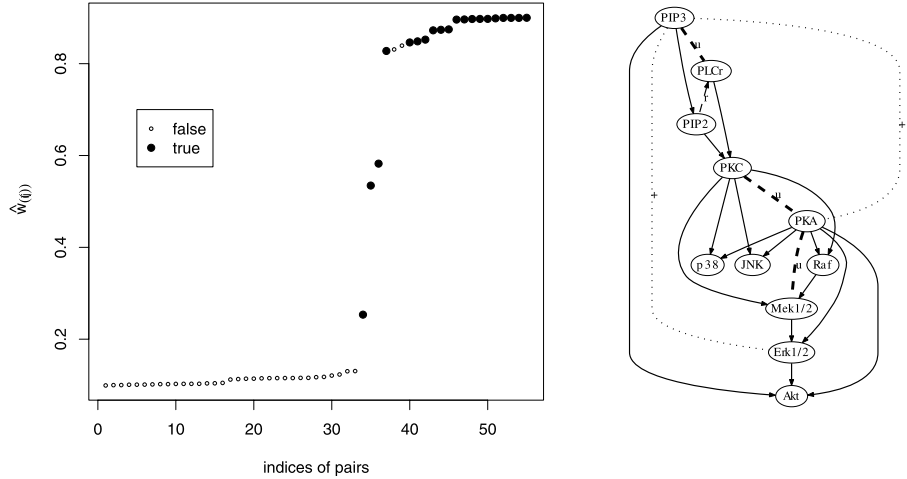}

\caption{Inference from RHM with $v_{ij}=0.1$ on the simulated data
with constant $\sigma_i^I$. Left: posterior means $\hat{w}_{(i,j)}$ of
$w_{ij}$, sorted in increasing order. Right: inferred networks with
$u_1'=0.2$, $u_3 \in(0.3,0.5)$. Solid arrowed lines represent
correctly inferred true edges, dashed thick lines with labels ``u''
represent edges whose directions cannot be determined from the
simulations, dashed arrowed thin lines with labels ``r'' represent
reversed edges, and dotted lines with labels ``+'' represent false
positive edges.}\label{wMean-rh.fig}
\end{figure}

When we have multiple MCMC runs, we infer the causal relation of each
edge based on the majority vote of the directions inferred from each
MCMC run.
In fact, these five runs lead to almost identical causal inference for
the common associations [based on $u_1 \in(0.3,0.4)$] when we take
$u_2=0.1$ and $u_3 \in(0.3,0.5)$.
The choices of $u_2$ and $u_3$ are affected by the value of $v_{ij}^{(k)}$.
Choosing $v_{ij}^{(k)} \equiv0.1$ ensures that most $\hat
{w}_{ij}^{(k)}$ are either above 0.9 or below 0.1. The streams with
$\hat{w}_{ij}^{(k)} \le0.1$ for all $k$ contain too weak a signal to
provide sufficient information for causal inference. So we take $u_2=0.1$.
The small value of $v_{ij}^{(k)}$ also leads to a great difference in
$\hat{w}_{ij}^{(k)}$ for different experiments when protein $i$ or $j$
is intervened. In this simulation study, intervention of the child node
for one edge leads to a decrease of at least 0.5 in $\hat{w}_{ij}^{(k)}$.
Any value of $u_3$ in (0.3, 0.5) leads to the same directional
inference for the inferred associations.

\subsubsection{Inference from RHM}\label{sec3.1.2}
RHM requires that $w_{ij}=w_{ji}$ and $w^{(k)}_{ij}=w^{(k)}_{ji}$ for
all $i$, $j$, and $k$.
This restriction aims at avoiding the nonconsistent directional
inferences based on $w^{(k)}_{ji}$ and $w^{(k)}_{ij}$ separately as HM does.
We plot the posterior means $\hat{w}_{(i,j)}$ of $w_{ij}$ in Figure
\ref{wMean-rh.fig} where we take $v_{ij}=0.1$.
Similar to Figures \ref{wMean.fig} and~S1, true associations tend to
have higher values of $\hat{w}_{(ij)}$.
Setting $u'_1=0.2$, we infer 22 associations with 2 false positives.
Applying the criteria listed in Supplementary Material S3 [\citet
{supp}], we infer the causal network as shown in Figure \ref{wMean-rh.fig}.
Compared to the network in Figure \ref{sml18-network.fig}, RHM leads
to a network with 16 true directed edges, 3~edges whose directions are
undetermined, 1 reversed edge, and 2 false edges when we take
$u'_1=0.2$ and $u_3 \in(0.3,0.5)$.
If we increase the threshold $u'_1$ to a value where there is a big
jump, for example, 0.3, we will miss 1 true directed edge.

When a bigger value $v_{ij}=10$ is applied, the differences of the
posterior means $\hat{w}_{(i,j)}$ of $w_{ij}$ become much smaller
between the true and false associations (Figure~S2 in the Supplementary
Material [\citet{supp}]). This together with the fact that bigger
values of $v_{ij}$ lead to smaller changes in experimental level
probabilities results in our conclusion that a small $v_{ij}$ is
preferred for causal network inference.

\subsubsection{Inference from NHM}
\begin{figure}

\includegraphics{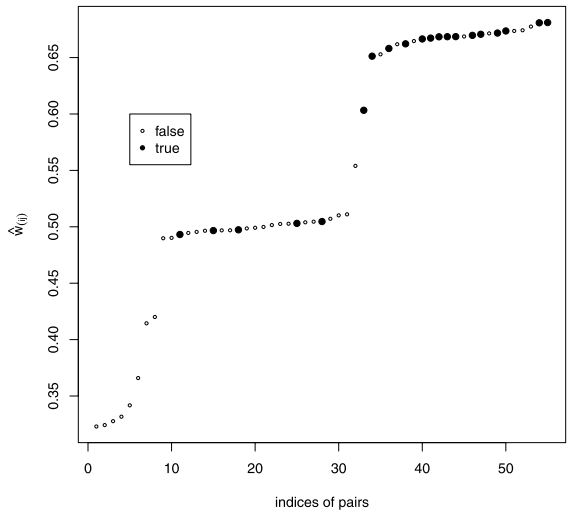}

\caption{Posterior means $\hat{w}_{(i,j)}$ of
$(w_{ij}+w_{ji})/2$, sorted in increasing order, from one MCMC run of
NHM. Small empty and large solid circles represent the false and true
associations, respectively.}\label{wMean-nonH.fig}
\end{figure}

Ignoring the effect of perturbations on signaling pathway,
NHM assumes a common coefficient $\alpha_{ij}$ in the linear
regression models across all experimental conditions.
From this model, we can only infer whether there is an association
between two proteins.
Similar to the inferred posterior means from HM, $\hat{w}_{(i,j)}$
from NHM also tend to take higher values for true associations (Figure
\ref{wMean-nonH.fig}), but with two differences.
First, the range of $\hat{w}_{(i,j)}$ from NHM is smaller.
In other words, compared to HM, NHM leads to smaller values of the
biggest $\hat{w}_{(i,j)}$, and larger values of the smallest $\hat
{w}_{(i,j)}$.
So the support for true associations and the evidence against false
associations are weaker.
Second, the dominance of high values of true associations is not as
strong as that from HM.
More false associations take higher values of $\hat{w}_{(i,j)}$ than
the hierarchical inference.
If we take 0.6 as a~threshold, we infer 23 associations with 15 true
and 8 false.
Taking 0.45 as the threshold, we recover all the true associations, but
27 false ones are also inferred.
More importantly, we cannot determine the directions of associations
from NHM because perturbation information is not utilized in this
model.\looseness=-1

\subsection{\texorpdfstring{Variable intrinsic variances $(\sigma^I_i)^2$}{Variable intrinsic variances}}\label{sec3.2}
\begin{figure}[b]

\includegraphics{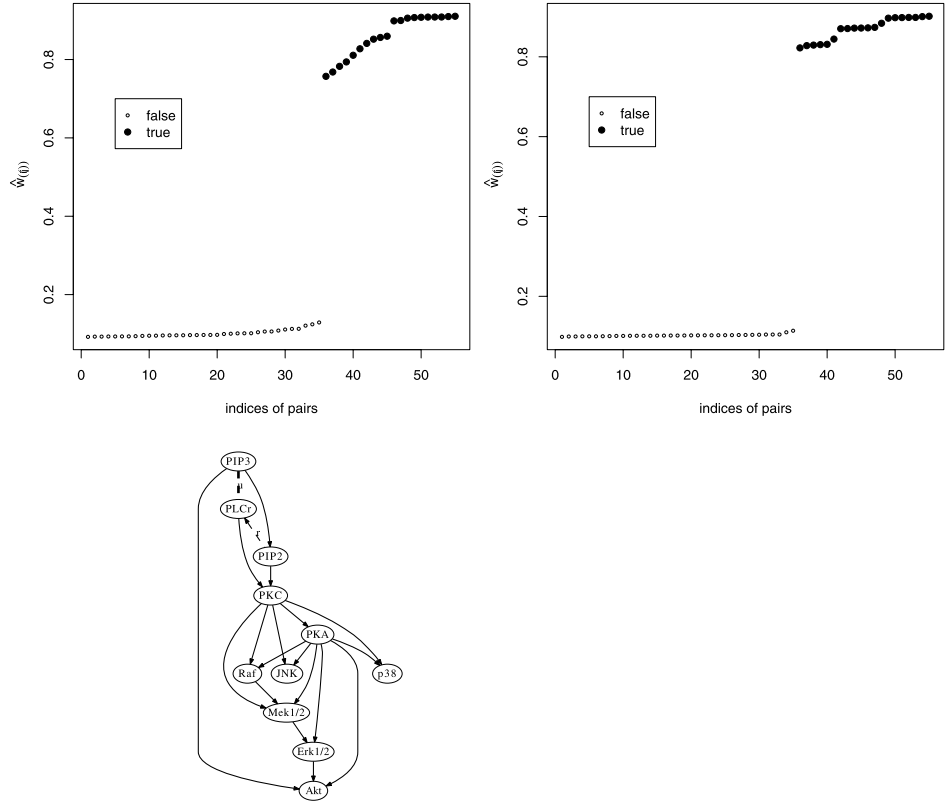}

\caption{Inference results for the simulated data with variable
intrinsic variances. Upper panel: posterior means $\hat{w}_{(i,j)}$ of
$w_{ij}$, sorted in increasing order from HM \textup{(left)} and RHM with
$v_{ij}=0.1$ \textup{(right)}. Lower panel: inferred networks from both models
with $u_1 \in(0.2,0.7)$ and $u_3=0.3$. Solid arrowed lines represent
correctly inferred true edges, dashed thick lines with labels ``u''
represent edges whose directions cannot be determined from the
simulations, and dashed arrowed thin lines with labels ``r'' represent
reversed edges.}\label{sml34.fig}
\end{figure}

We then consider the case where variances of intrinsic noises vary for
different proteins.
In this case, both HM and RHM with $v_{ij}=0.1$ clearly separate the
true associations from the false ones in the plot of the posterior
means $\hat{w}_{(i,j)}$ (Figure \ref{sml34.fig}). The causal networks
inferred from both models are the same, with 18 correctly inferred true
edges, 1 reversed edge, and 1 edge whose direction is undetermined
[$u_1 \in(0.2,0.7)$, $u_2=0.1,$ and $u_3=0.3$].
As in Section \ref{sec3.1.2}, RHM with a~bigger value $v_{ij}=10$ leads to
association inference with bigger false positive rate and smaller
changes in experimental level probabilities when a child node is
perturbed (Figure S2 in the Supplementary Material [\citet
{supp}]). NHM is not applied here and thereafter since it does not
provide causal relations.

\begin{table}
\caption{Summary of pathway inference in simulation study}\label{net.tab}
\begin{tabular*}{\tablewidth}{@{\extracolsep{\fill}}lccccccc@{}}
\hline
\textbf{Data}&\textbf{Methods}& \textbf{True} & \textbf{Undetermined} & \textbf{Reversed} & \textbf{Missing} & \textbf{False} &
\textbf{Hamming}\\
&&&&&&&\textbf{distance}\\
\hline
Data-1 & HM & 14 & 5 & 1 & 0 & 2 & \hphantom{0}8\\
& RHM & 16 & 3 & 0 & 1 & 2 & \hphantom{0}6\\[3pt]
Data-2 & HM & 18 & 1 & 1 & 0 & 0 & \hphantom{0}2\\
& RHM & 18 & 1 & 1 & 0 & 0 & \hphantom{0}2\\[3pt]
Data-$1^t$ & HM & \hphantom{0}9 & 4 & 1 & 6 & 4 & 15\\[3pt] 
Data-$2^t$ & HM & 14 & 1 & 0 & 5 & 6 & 12\\ 
\hline
\end{tabular*}
\legend{All are based on $u_1=0.4$ and $u_3=0.3$.
The hamming distance is the minimum number of simple operations needed
to go from the inferred graph to the true graph. Here simple operations
include adding or removing an edge, and adding, removing, or changing
the direction of an edge.
Data-1: simulated data in Section \ref{sec3.1} with constant intrinsic
variances. Data-2: simulated data in Section~\ref{sec3.2} with varying intrinsic
variances. Data-$1^t$: simulated data with parameter settings in Data-1
and intrinsic noises sampled from $t(1)$. Data-$2^t$: simulated data
with parameter settings in Data-2 and intrinsic noises sampled from
$t(1)$.}
\end{table}

\subsection{Heavy tail distribution for intrinsic noise}\label{sec3.3}
Considering the possibility of nonnormality for real biological
processes, we simulate data where the expression levels of proteins
have heavy tail distribution. This is realized by simulating $\epsilon
_{\mathit{ink}}^I \sim\mathrm{t}(1)$ for each protein $i$ under each
experimental condition~$k$.
We reuse the parameter settings in Sections \ref{sec3.1} and \ref{sec3.2} so that the
performance of our methods on the normal and nonnormal cases can be
easily compared.
We summarize the network inference results in Table \ref{net.tab}. Due
to the model misspecification when we use HM to analyze these heavy
tail distributed data, we infer networks with more false positive and
false negative edges. Therefore, our current model needs to be extended
to analyze heavy tail data.

\section{Case study}\label{sec4}
The Mitogen-Activated Protein Kinase (MAPK) pathways transduce a large
variety of external signals, leading to a wide range of cellular
responses such as growth, differentiation, inflammation, and apoptosis.
External stimuli are sensed by cell surface markers, then travel
through a~cascade of protein modifications of signaling proteins, and
eventually lead to changes in nuclear transcription.
Single cell interventional data of 11 well-studied proteins from the
MAPK pathways were originally generated by \citet{sachs-etal-05} using
the intracellular multicolor flow cytometry technique.
This pathway was perturbed by 9 different stimuli, each targeting a
different protein in the selected pathway (Figure~\ref{pathway.fig}
and Table \ref{condition}).
\citet{sachs-etal-05} applied Bayesian network analysis to infer the
causal protein-signaling network.
Correcting the bias in the commonly used algorithm proposed by \citet
{friedman-koller}, \citet{ellis-wong} reanalyzed this data set through
sampling BN structures from the correct posterior distribution.
Both studies used the discretized data where the protein expression
levels were grouped into three levels: ``low,'' ``middle,'' and ``high.''
The inhibited molecules were set at ``low'' values, and activated
molecules were set to level ``high.''
We apply our method to this data and compare the results with those
from \citet{ellis-wong}.

We infer the networks using HM and RHM with $v_{ij}=0.1$ and
$v_{ij}=10$. Each analysis has five MCMC runs.
Figure \ref{bde.fig} shows the inferred posterior means $\hat
{w}_{(i,j)}$ in one MCMC run (more can be found in Supplementary
Figures S4 $\sim$ S6 [\citet{supp}]), and the inferred networks
from five MCMC runs, from each method.
We use the same symbols as in simulation studies to indicate true or
false inferences, where the ``true'' network is taken to be the network
in Figure 3 of \citet{sachs-etal-05}, which is the current
understanding of this pathway.

Compared to HM, RHMs lead to fewer true associations with high values
of $\hat{w}_{(i,j)}$ ($v_{ij}=10$) or smaller gaps of $\hat
{w}_{(i,j)}$ between most true and false associations ($v_{ij}=0.1$).
Taking the threshold $u_1=0.2$ and requiring $u_f \ge0.6$ in five runs
of HM, we get 21 associations, with 5 missing edges and 6 false positives.
Requiring $u'_1=0.11$ and $u_f \ge0.6$ in RHM with $v_{ij}=0.1$, we
get 19 associations, with 5 missing edges and 4 false ones.
The threshold 0.11 exceeds the value (0.1) from the permutation study
by only a small amount, implying that RHM offers weaker support to true
associations than HM.
Setting $u'_1=0.994$ and $u_f \ge0.6$ in RHM with $v_{ij}=10$, we only
get 14 associations, with 8 missing and 2 false associations.

\begin{figure}

\includegraphics[scale=0.92]{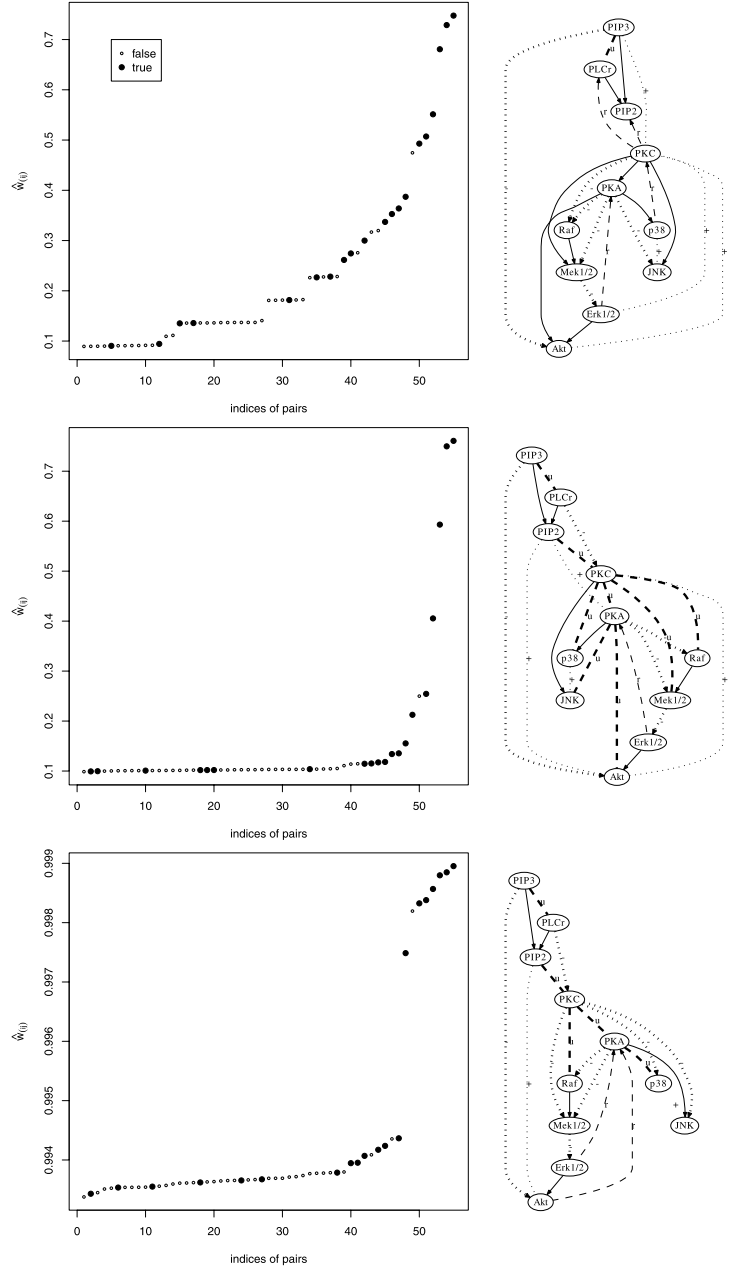}

\caption{Inference results for the real data. From top to bottom:
HM, RHM with $v_{ij}=0.1$, and RHM with $v_{ij}=10$. In networks,
solid arrowed lines represent correctly inferred true edges, dashed
thick lines with labels ``u'' represent edges whose directions
cannot be determined from the simulations, dashed arrowed thin
lines with labels ``r'' represent reversed edges, dotted arrowed
thick lines represent missing edges, and dotted thin lines with
labels ``+'' represent false positive edges.}\label{bde.fig}
\end{figure}

\begin{table}[b]
\caption{Summary of the inferred networks applying
different methods to the real data}\label{real.tab}
\begin{tabular}{@{}lcccccc@{}}
\hline
& \textbf{True} & \textbf{Undetermined} & \textbf{Reversed} & \textbf{Missing} & \textbf{False} & \textbf{Hamming distance}\\
\hline
HM & 9 & 1 & 4 & 6 & 4 & 15 \\
RHM $v_{ij}=0.1$ & 6 & 8 & 1 & 5 & 4 & 18 \\
RHM $v_{ij}=10$ & 5 & 5 & 2 & 8 & 2 & 17 \\
mHM & 6&5&1&8&4& 18 \\
BN & 9 & 0 & 3 & 8 & 6 & 17\\
\hline
\end{tabular}
\legend{Here mHM denotes the modified model
described in Supplementary Material S1 [\citet{supp}] which
models the varying variances of intrinsic noises.}
\end{table}

From RHM, we can only correctly infer the directions of five or six edges.
The causal relations for most inferred associations can not be determined.
But HM leads to a better result: 9 true directed edges, 1
direction-undetermined, 4 reversed, 6 missed, and 4 false edges, under
the thresholds $u_1=0.2$, $u_2=0.1$, $u_3=0.3$, and $u_f \ge0.8$
(Figure \ref{bde.fig} and Table \ref{real.tab}).
This inferred network is comparable with that from \citet{ellis-wong},
which contains 9 true directed edges, 3 reversed, 8 missed, and 6 false edges.
The Hamming distances of these two networks to Figure~\ref
{pathway.fig} are 15 and 17, respectively.
These results are summarized in Table \ref{real.tab}.

In MCMC analysis, we take $\beta_i=\gamma_j=1$ for $i=1,2$ and
$j=1,\ldots,6$.
To check the sensitivity of HM, we also consider other values: $\gamma
_i=0.1$ or 100, and $\beta_j=0.1$ or 0.0001.
Taking $\beta_j=0.1$, we get a network (not shown) with 9 true
directed edges, 1 direction-undetermined, 4 reversed, 6 missed, and 6
false associations.
Other values of the hyperparameters result in 1$\sim$3 fewer true
associations, and at least 3 fewer true directed edges.
All these results are based on 5,500,000 iterations of MCMC updates in
each run, which take about 20 hours on a node with an Intel(R) Xeon(R)
3 GHz CPU and a 16G memory.

%

\section{Discussion}\label{sec5}
We have proposed hierarchical statistical methods to infer a signaling
pathway from single cell data collected from a set of perturbation experiments.
The advantage of this method is that it provides a more explicit
framework to relate the activity levels of different proteins.
In our models, we assume that the activity level of each protein is
linearly associated with a small subset of other proteins under each condition.
Using a Bayesian hierarchical structure, we model the existence of an
association between two proteins both at the overall level and at the
experimental level. 
The overall-level probabilities measure the strength of associations
between any two proteins across all experiments.
The experimental-level probabilities reflect the changes of
associations between proteins under different conditions.
Our inferential procedure consists of two steps.
First we infer the existence of an association between any pair of
proteins based on the overall-level probabilities.
Then for those pairs of proteins inferred to be associated, we infer
the directions of the causal relations based on the changes in the
experimental level probabilities.
The basic rationale in our causal inference is that for two associated
proteins, controlling over the target molecule destroys the
association, while perturbing the regulatory molecule does not.

We consider hierarchical models with (RHM) and without (HM) the
restriction that $w_{ij}=w_{ji}$ and $w_{ij}^{(k)}=w_{ji}^{(k)}$ for
each $k$.
For RHM, we have to specify the hyperparameter $v_{ij}$ prior to MCMC analysis.
We have considered the inference results when the value of $v_{ij}$ is
set at 0.1 and 10.
Higher values of $v_{ij}$ lead to higher ranges of the inferred
overall-level and experimental-level probabilities, and smaller changes
in experimental-level probabilities.
In HM, the experimental-level probabilities can be integrated out, so
the posterior inference of other parameters is independent of $v_{ij}$.
Hence, the choice of associations, which is based on the overall-level
probabilities, is independent of $v_{ij}$.
We only need to specify $v_{ij}$ in the causal inference.
To better reflect the changes of the experimental-level probabilities,
we suggest smaller values for $v_{ij}$, for example, $v_{ij}=0.1$.
Both HM and RHM perform well in simulation studies.

We need to choose thresholds to infer the causal network: $u_1$ for
association inference and $u_2$ and $u_3$ for causal directional inference.
Noting the jumps in the plots of $\hat{w}_{(ij)}$, we propose to
choose the threshold $u_1$ where there are great differences in sorted
$\hat{w}_{(ij)}$.
This is easily determined when variations in the data are well captured
by the proposed hierarchical models (e.g., Figures \ref{wMean.fig} and
\ref{sml34.fig}). If there are no great differences in the sorted
overall-level probabilities (e.g., Figure \ref{bde.fig}), one may
decide the number of edges to be included and then choose the top ones.
Threshold $u_2$, which is taken as 0.1 in our study, can be chosen
based on the experimental-level probabilities of those unassociated
pairs of proteins.
Threshold $u_3$ is closely related to $v_{ij}$, which measures the
variability in $\hat{w}_{ij}^{(k)}$ in that a smaller $v_{ij}$ leads
to greater variabilities in $\hat{w}_{ij}^{(k)}$ between the
experiments when the target protein is and is not intervened.
When $v_{ij}=0.1$, a difference of 0.3 in $\hat{w}_{ij}^{(k)}$ is
enough to show the effect of intervening the target protein.

Compared to the nonhierarchical model, hierarchical models have at
least two advantages.
First, the hierarchical structure allows information borrowing across
different experiments while allowing for differences among experiments,
leading to a more clear-cut inference on whether two proteins are related.
Second, this modeling framework allows us to infer causal relationships
between proteins from the presence and absence of the association
across different perturbation conditions.
Overall, our proposed hierarchical modeling provides a general
framework for inferring networks from high-throughout data.

There are several possible ways of extending this model. In
Supplementary Material S1 [\citet{supp}] we modify HM by
incorporating varying variances of intrinsic noises under different
experimental conditions. The modified model does not outperform HM in
our simulation study. It is interesting to investigate when the varying
variances of intrinsic noises are not ignorable and incorporating them
improves the network inference.
We also find in our simulation study that applying our methods to data
where intrinsic noises are sampled from heavy tail distributions
results in power loss in pathway inference. Therefore, there is a need
to extend this hierarchical structure to model nonnormal data.

\begin{supplement}
\stitle{Additional descriptions and results of
hierarchical models\\}
\slink[doi]{10.1214/10-AOAS425SUPP} 
\slink[url]{http://lib.stat.cmu.edu/aoas/425/supplement.pdf}
\sdatatype{.pdf}
\sdescription{Materials include description and simulation results of
the hierarchical model (mHM) with varying variances of intrinsic noises
$(\sigma_{ik}^I)^2$, MCMC algorithm for the hierarchical model (HM),
direction inference for the restricted hierarchical model (RHM), and
additional figures of posterior inference and networks.}
\end{supplement}


\printaddresses


\begin{thebibliography}{99}

\bibitem[\protect\citeauthoryear{Dobra et~al.}{2004}]{dobra-etal04}
\textsc{Dobra}, A., \textsc{Hans}, C., \textsc{Jones}, B., \textsc
{Nevins}, J., \textsc{Yao}, G. and \textsc{West}, M. (2004). Sparse
graphical models for exploring gene expression data.
\textit{J. Multivariate Anal.} \textbf{90} 196--212.
\MR{2064941}

\bibitem[\protect\citeauthoryear{Ellis and Wong}{2008}]{ellis-wong}
\textsc{Ellis}, B. and \textsc{Wong}, W.~H. (2008). Learning causal
Bayesian network structures
from experimental data.
\textit{J. Amer. Statist. Assoc.} \textbf{103}
778--789.
\MR{2524009}

\bibitem[\protect\citeauthoryear{Friedman and Killer}{2003}]{friedman-koller}
\textsc{Friedman}, N. and \textsc{Killer}, D. (2003). Being Bayesian
about network structure.
\textit{Machine Learning} \textbf{50} 95--126.

\bibitem[\protect\citeauthoryear{Heckerman et~al.}{2000}]{heckerman-etal-00}
\textsc{Heckerman}, D., \textsc{Chickering}, D.~M., \textsc{Meek},
C., \textsc{Rounthwaite}, R. and \textsc{Kadie}, C.
(2000). Dependency networks for inference, collaborative filtering, and data
visulization.
\textit{J.~Mach. Learn. Res.} \textbf{1} 49--75.

\bibitem[\protect\citeauthoryear{Herzenberg et~al.}{2002}]{flow02}
\textsc{Herzenberg}, L.~A., \textsc{Parks}, D., \textsc{Sahaf}, B.,
\textsc{Perez}, O., \textsc{Roederer}, M. and
\textsc{Herzenberg}, L.~A. (2002). The history and future of the
fluorescence activated
cell sorter and flow cytometry: A~view from Stanford.
\textit{Clinical Chemistry} \textbf{48} 1819--1827.


\bibitem[\protect\citeauthoryear{Lauritzen}{1996}]{lauritzen}
\textsc{Lauritzen}, S.~L. (1996). \textit{Graphical Models}.
Clarendon Press, Oxford.
\MR{1419991}

\bibitem[\protect\citeauthoryear{Liu and Ringn\'
{e}r}{2007}]{Liu-Ringner-2007}
\textsc{Liu}, Y. and \textsc{Ringn\'{e}r}, M. (2007). Revealing
signaling pathway deregulation by
using gene expression signatures and regulatory motif analysis.
\textit{Genome Biology} \textbf{8} R77.1--R77.10.

\bibitem[\protect\citeauthoryear{Luo and Zhao}{2010}]{supp}
\textsc{Luo}, R. and \textsc{Zhao}, H. (2010). Supplementary material
for ``Bayesian
hierarchical modeling for signaling pathway inference from single cell
interventional data.'' DOI: \href
{http://dx.doi.org/10.1214/10-AOAS425SUPP}{10.1214/10-AOAS425SUPP}.

\bibitem[\protect\citeauthoryear{Pe'er}{2005}]{peer05}
\textsc{Pe'er}, D. (2005). Bayesian network analysis of signaling
networks: A primer.
\textit{Science's STKE} \textbf{281} 1--12.

\bibitem[\protect\citeauthoryear{Pe'er et~al.}{2001}]{peer01}
\textsc{Pe'er}, D., \textsc{Regev}, A., \textsc{Elidan}, G. and
\textsc{Friedman}, N. (2001). Inferring subnetworks
from perturbed expression profiles.
\textit{Bioinformatics} \textbf{17 Suppl.} S215--S224.

\bibitem[\protect\citeauthoryear{Perez and Nolan}{2002}]{nolan02}
\textsc{Perez}, O.~D. and \textsc{Nolan}, G. (2002). Simultaneous
measurement of multiple active
kinase states using polychromatic flow cytometry.
\textit{Nature Biotechnology} \textbf{20} 155--162.

\bibitem[\protect\citeauthoryear{Sachs et~al.}{2005}]{sachs-etal-05}
\textsc{Sachs}, K., \textsc{Perez}, O., \textsc{Pe`er}, D., \textsc
{Lauffenburger}, D.~A. and \textsc{Nolan}, G.~P.
(2005).
Causal protein-signaling networks derived from multiparameter single-cell
data.
\textit{Science} \textbf{308} 523--529.

\bibitem[\protect\citeauthoryear{Sch\"{a}fer and
Strimmer}{2005}]{schafer-strimmer05}
\textsc{Sch\"{a}fer}, J. and \textsc{Strimmer}, K. (2005). An
empirical {B}ayes approach to
inferring large-scale gene association networks.
\textit{Bioinformatics} \textbf{21} 754--764.

\bibitem[\protect\citeauthoryear{Wei and Li}{2007}]{li07}
\textsc{Wei}, W. and \textsc{Li}, H. (2007). A Markov random field
model for network-based
analysis of genomic data.
\textit{Bioinformatics} \textbf{23} 1537--1544.

\bibitem[\protect\citeauthoryear{Wei and Li}{2008}]{li08}
\textsc{Wei}, W. and \textsc{Li}, H. (2008). A hidden
spatial--temporal Markov
random field model for network-based analysis of time course gene
expression data.
\textit{Ann. Appl. Statist.} \textbf{2} 408--429.
\MR{2415609}

\bibitem[\protect\citeauthoryear{Werhli, Grzegorczyk and
Husmeier}{2006}]{werhli-etal-06}
\textsc{Werhli}, A.~V., \textsc{Grzegorczyk}, M. and \textsc
{Husmeier}, D. (2006). Comparative evaluation
of reverse engineering gene regulatory networks with relevance networks,
graphical Gaussian models and Bayesian networks.
\textit{Systems Biology} \textbf{22} 2523--2531.

\end{thebibliography}
\end{document}